\documentclass[twocolumn, pra]{revtex4-1}
\usepackage{amsmath}
\usepackage{graphicx}
\usepackage{amssymb}
\usepackage{latexsym}
\usepackage{verbatim,times,bbm}
\usepackage{latexsym}

\newcommand{\bra}[1]{\langle #1 |}
\newcommand{\ket}[1]{| #1 \rangle}

\begin{document}

\title{A Quantum Enigma Machine: Experimentally Demonstrating Quantum Data Locking$^{\star}$}

\author{Daniel J. Lum}
\affiliation{Department of Physics and Astronomy, University of Rochester, Rochester, New York 14627, USA}
\altaffiliation{Center for Coherence and Quantum Optics, University of Rochester, Rochester, New York 14627, USA}
\email[]{daniel.lum@rochester.edu}
\author{M. S. Allman}
\affiliation{National Institute of Standards and Technology, 325 Broadway, Boulder, Colorado 80305, USA}
\author{Thomas Gerrits}
\affiliation{National Institute of Standards and Technology, 325 Broadway, Boulder, Colorado 80305, USA}
\author{Cosmo Lupo}
\affiliation{Research Laboratory of Electronics, Massachusetts Institute of Technology, Cambridge, MA 02139, USA}
\author{Varun B. Verma}
\affiliation{National Institute of Standards and Technology, 325 Broadway, Boulder, Colorado 80305, USA}
\author{Seth Lloyd}
\affiliation{Department of Mechanical Engineering, Massachusetts Institute of Technology, Cambridge, MA 02139, USA}
\author{Sae Woo Nam}
\affiliation{National Institute of Standards and Technology, 325 Broadway, Boulder, Colorado 80305, USA}
\author{John C. Howell}
\affiliation{Department of Physics and Astronomy, University of Rochester, Rochester, New York 14627, USA}
\altaffiliation{Center for Coherence and Quantum Optics, University of Rochester, Rochester, New York 14627, USA}

\date{\today}



\begin{abstract}
Claude Shannon proved in 1949 that information-theoretic-secure encryption is possible if the encryption key is used only once, is random, and is at least as long as the message itself. Notwithstanding, when information is encoded in a quantum system, the phenomenon of quantum data locking allows one to encrypt a message with a shorter key and still provide information-theoretic security. We present one of the first feasible experimental demonstrations of quantum data locking for direct communication and propose a scheme for a quantum enigma machine that encrypts 6 bits per photon (containing messages, new encryption keys, and forward error correction bits) with less than 6 bits per photon of encryption key while remaining information-theoretically secure.

\end{abstract}

\maketitle

\section{Introduction}

One of the fundamental results of classical information theory, due to Claude Shannon, is that the secure 
encryption of a message against an adversary with infinite computing power, i.e. information-theoretic security, requires the single use of a random secret key at least of equal length \cite{shannon1949communication}.
Therefore, in the classical setting, the use of a shorter key may jeapordize the security 
of a communication protocol.
Surprisingly, Shannon's stringent criterion can be circumvented if information is encoded in a quantum system.
According to a phenomenon known as quantum data locking (QDL), a short secret key can encrypt a much
longer message if the latter is encoded in a quantum system, in such a way that
the encryption is provably secure against an eavesdropper that intercepts and measures the quantum system
\cite{divincenzo2004locking,HLSW04,DFHL10,Fawzi2013}.

Together with other uniquely quantum phenomena, such as Bell inequalities and teleportation, 
QDL represents one of the strongest violations of classical information
theory in the quantum setting. 
Quantum information theory shows that some strong QDL protocols allow for information-theoretic security 
while encrypting a message with a key that is \emph{exponentially} shorter than the message itself. 
Here we present one of the first experimental realizations of QDL as a ``quantum enigma machine'' \cite{lloyd2013quantum,GuhaPhysRevX},
i.e., an optical implementation of QDL, allowing secure direct communication under the assumption that an adversary must periodically make 
measurements on the intercepted quantum state. During the preparation of this manuscript, we became aware of an alternate experiment performed at approximately the same time as ours \cite{liu2016experimental} that limits an eavesdropper's accessible information to half that of the legitimate receiver by utilizing the protocol found in \cite{divincenzo2004locking}. Our experimental realization, which utilizes higher dimensions in free-space propagation, allows us to limit an eavesdropper's accessible information to an arbitrarily small amount. This allows us, in principle, 
to securely transfer 6 bits per photon of message, new secret key, and forward error correction
via direct secret communication while encrypting each photon with a
key of strictly less than 6 bits.

\section{Quantum Data Locking}

In a QDL protocol, outlined in \cite{divincenzo2004locking,HLSW04,DFHL10,Fawzi2013,lupo2015quantumArXiv}, Alice attempts to transmit messages encoded onto quantum states. Alice first locks the data to be transmitted by applying a random unitary operation on her quantum state, and finally sends the quantum state 
through a public channel to Bob. Alice and Bob share a private key describing which unitary operations Alice used such that Bob can accurately perform the inverse unitary 
operations and unlock the original message. The objective, for what concerns the security of the protocol, is to guarantee that an eavesdropper, Eve, cannot retrieve the original message if she intercepts the quantum 
state without access to the private key. This security is provided by limiting Eve's accessible information, being defined as the maximum classical mutual information between the message Alice is trying to send and an optimal measurement performed by Eve \cite{divincenzo2004locking}.

The security of QDL is granted under proper assumptions. While we allow Eve to have unbounded computational power, we must assume she has imperfect quantum technology. In particular, we must assume Eve has a quantum memory with finite storage time. This assumption compensates for the fact that, in general, the accessible
information is not composable \cite{PhysRevLett.98.140502}. 
Composable security asserts that the security of individual communication protocols is preserved when used as subroutines within an overarching communication protocol \cite{canetti2001foundations}. Note that the same constraints can also be imposed on Bob, meaning QDL does \emph{not} require Bob to have better quantum technology than Eve.
The assumptions of QDL are more stringent than those of standard 
Quantum Key Distribution (QKD) and are analogous to QKD systems utilizing a Bounded Storage Model which secures a key with the assumption that Eve has a finite amount of memory, whether classical or quantum \cite{damgaard2007tight}. QDL requires that Eve has a quantum memory with finite storage time along with the use of a pre-shared secret key. However, theory states that QDL can tolerate higher loss. For example, in the case presented in \cite{lupo2015quantumArXiv}, QDL can tolerate up to 66\% channel loss, as opposed to 50\% loss in the most robust QKD protocols, while also promising unprecedented high rates \cite{GuhaPhysRevX,lupo2015quantumArXiv,lupo2014PhysRevLett,PhysRevA.92.062312}.

In a typical QDL protocol, Alice and Bob first decide on the number of possible messages $M$ that will be encoded into $n$ quantum states ($n$ single photons in our case) and transmitted over $n$ uses of a 
quantum channel. Each message will then be $\mathrm{log}_2(M)$ bits in length. Next, they then privately agree on a $\mathrm{log}_2(K)$ bit key that specifies a line within a public code book containing $K$ unique lines. Each line of the code book specifies the sequence of $n$ unitary 
operations on the $n$ quantum states that Alice sends to Bob. Specifically, each line lists a sequence of $n$ random seeds that generate a pseudo-random unitary transform. Alice applies these unitary transforms to the $n$ quantum states and sends them to Bob. The total number of all possible enciphered states is $N = MK$. For efficient QDL, we require $K \ll M$ \cite{lupo2014PhysRevLett,lupo2014robust}. If Eve knew precisely which of the $N$ 
possible enciphered states was sent, she could prepare $K$ copies of the full transmission and perform the inverse of the $K$ unitary transformation sequences on 
identical copies of the enciphered state to determine which message was sent. However, Eve cannot identify the original message from the set of possible $N$ states with only $n \ll N$ measurements, assuring the security of this protocol.

Under proper scenarios, QDL has the potential to replace some QKD systems and may serve as a better option to alternate Quantum-Secure Direct Communication (QSDC) protocols. QDL has, in principle, the possibility of being a valuable alternative to standard QKD because QDL boasts a higher secure-key rate (per photon) when the limited quantum memory of an eavesdropper becomes a practical assumption \cite{lupo2014PhysRevLett}. Of course, the actual field-tested secure bit rate, based upon technological limitations and channel losses, will be the final deciding factor as to whether QDL or QKD is used. It should also be noted that QDL holds many similarities with QSDC, whose goal is to transmit information directly over a quantum channel without having to first establish a shared private key, if at all. The core of QSDC security relies on fundamental quantum principles such as the no-cloning theorem \cite{wootters1982single}, uncertainty principles, or quantum correlations. QSDC protocols ensure security through the ability to either detect an eavesdropper before information is leaked \cite{PhysRevA.69.052319,PhysRevA.71.044305} or to prevent information leakage by denying outsiders access to the entirety of a correlated quantum state \cite{Schumacher:07,zheng2014quantum,lee2006quantum,farouk2015generalized}. In many cases, these protocols either require that a single quantum state is transmitted over a quantum channel twice (from Alice, out to Bob, and then back) or necessitate the distribution of a correlated, possibly entangled, quantum state. It is clear that QDL depends on the initial establishment of a short secret key. However, none of the alternate QSDC protocols appear to utilize the locking effect afforded through the definition of the accessible information. In addition, because of the experimental difficulty with distributing entangled states or with losses from having a state traverse a quantum channel twice, no experimental demonstrations have been published in a peer-reviewed journal to date.

\section{Quantum enigma}

\begin{figure*}
\centering
\includegraphics[width=.9\textwidth]{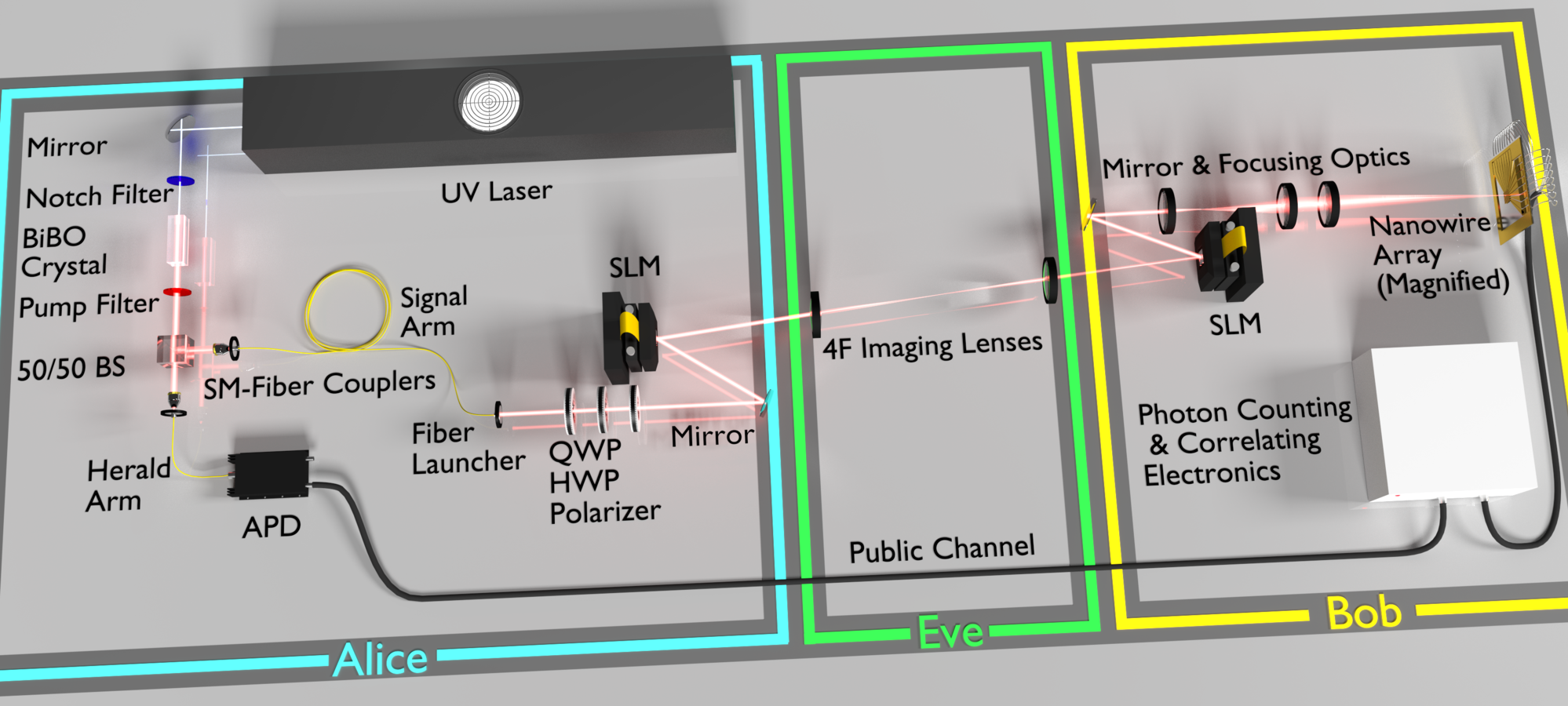}
\caption{\textbf{Experimental Diagram: } Ultraviolet (UV) laser light passes through a nonlinear bismuth 
triborate (BiBO) crystal to produce down-converted signal-herald photon pairs. After separating the photons with a beamsplitter (BS), Alice uses single-mode (SM) fibers to transfer the herald photon to an Avalance Photo Diode (APD) detector while sending the signal photon to be polarization prepped by a Quarter-Wave Plate (QWP), a Half-Wave Plate (HWP), and a polarizer for use with a Spatial Light Modulator (SLM). Alice then phase-modulates her signal photon with an SLM to encode and lock information. Imaging lenses map Alice's wavefront to Bob's SLM where he applies an inverse unitary 
phase operation to unlock the information before focusing his light onto a low-noise cryogenically cooled $8\!\times\! 8$ pixel nanowire array. Bob then uses the heralded detection event sent by Alice 
to herald the arrival of single photons. See the appendix for details.}
\label{fig:exp_setup}
\end{figure*} 

In a proof-of-principle experimental demonstration of QDL presented in Fig. \ref{fig:exp_setup}, we focus on one possible application for QSDC and encode 6 bits of information onto a single photon while encrypting 
that information with less than 6 bits of key. Single photon pairs are generated from the process of degenerate Spontaneous Parametric Down Conversion (SPDC) \cite{PhysRevA.31.2409} whereby a 
pump-laser photon at 325 nm down-converts into two 650 nm daughter photons, referred to here as signal and herald. The herald photon, detected by Alice's Avalanche Photodiode Detector (APD), is used to herald the presence of the 
signal photon on Bob's detector. Alice uses a $512\!\times\! 512$ pixel Spatial Light Modulator (SLM) to both 
encode information in the transverse linear phase of the signal photon's wavefront while also operating with a scrambling unitary operation specified by one of $n$ random number generating seeds in each line of the code book. Each scrambling unitary is a $128 \!\times\! 
128$ superpixel random binary phase mask composed of 0 and $\pi$ relative phase shifts. Utilizing properties of Fourier optics, a linear phase shift on a wavefront 
corresponds to a linear shift in the focal point of that wavefront's Fraunhofer diffraction pattern \cite{goodman2005introduction} while a scrambled wavefront has no 
well-defined focal point. Alice adds 1 of 64 linear phase patterns to a scrambling phase pattern and uses the resulting pattern to phase modulate her single photon with 
an SLM. She then transmits this photon to Bob. Once Bob receives the photon, he applies an inverse scrambling phase unitary with his SLM and focuses the resulting state 
onto a high-efficiency $8\!\times\! 8$ single-photon-detecting nanowire array \cite{allman2015near}. This detector array is cryogenically cooled to 0.8 K and was used because it exhibits one of the lowest dark count rates of any single-photon-detecting array to date (an essential criterion for accurately transmitting messages). If Bob unscrambles the wavefront properly, the photon will register on a detector 
pixel of Alice's choosing. Hence, linear phases encoded up to 
6 bits of encrypted information per photon.

\begin{figure}[h]
\centering
\includegraphics[width=.4\textwidth]{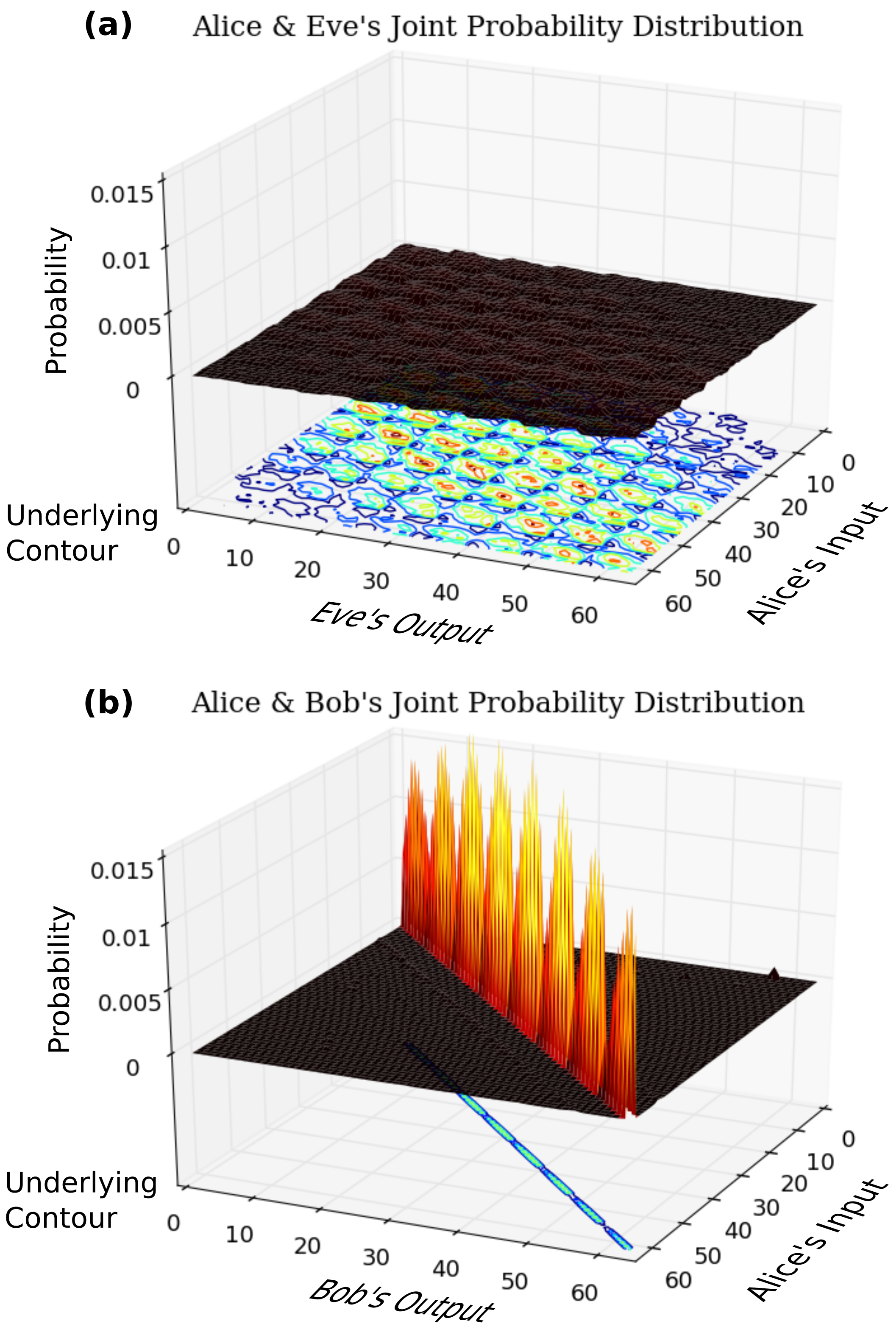}
\caption{\textbf{Joint Probability Distributions: }The joint probability distributions presented are derived 
from experimental data where Alice scans through 64 messages while locking the information with $128\!\times\! 128$ pixel binary phase masks. \textbf{a}, Eve randomly guesses at binary phase masks in hopes of unlocking the information while being allowed, in a worst-case scenario, to use Bob's properly aligned detector. \textbf{b}, Bob unlocks the messages with binary phase masks prescribed according to a secret key. The distribution with the highest mutual information is a normalized identity matrix with diagonal elements equal to $1/64 \approx .016$. The comb structure seen in (b) is due to a gradient of error rates across the detector array, possibly due to a slight misalignment of a lenslet array placed above the detector array. In summary, Alice and Eve's distribution is highly uncorrelated while Alice and Bob's distribution is highly correlated attesting the effectiveness of this locking method.}
\label{fig:jpd} 
\end{figure}

The effectiveness of the scrambling unitaries is depicted in 
Fig. \ref{fig:jpd}. Alice encrypted each of the 64 settings over 600 times while Bob used the secret key to unscramble the wavefront. Eve's optimal distribution was obtained in a best-case scenario using 
Bob's properly aligned SLM and detector while lacking the secret key. Therefore, she was forced to randomly guess inverse unitaries in an 
attempt to maximize her mutual information. 
The result of Bob's measurements are highly correlated with Alice's intended messages. Alternatively, Eve's measurements follow a highly-uncorrelated flat distribution for the probability of photodetection.

Within the appendix, we derive the necessary key rates needed to secure this system according to the accessible information. The experimental scheme realizes the theoretical proposal presented in \cite{lupo2015quantumArXiv}. A crucial parameter to determine the secret key length is the dimension of the quantum communication channel $d$. Roughly speaking, the setting with the most efficient use of key is when $d$ is comparable to the number of possible messages per photon. Our detector limits the set of possible messages per photon to 64. We utilize free-space propagation and present two potential ways to estimate $d$. A practical estimate is to assume that the eavesdropper is also constrained to measure with an $8\!\times\!8$ detector array identical to Bob's. In this way, the effective dimension is not greater than $d = 64$. An alternate, more conservative, estimate is to consider the transverse beam profile within the plane of the detector and evaluate the total number of unique (i.e., non-overlapping) spatial modes that may exist according to the propagation characteristics of the system. We use the practical estimate, $d = 64$, here within the main article. See the appendix for details along with alternate results for the conservative estimate.

\section{Secret key \& message transmission with forward error correction}

In order to reliably transmit messages and new secret keys, error rates must be low enough such that the data packets can be reliably transmitted and decrypted by Bob. Practically, these 
errors will never be omitted entirely. The overall error rate between what Alice transmits and what Bob receives is approximately $10\%$, originating from a combination of 
sources including SPDC photons scattering from the SLM's and background photons. However, this error rate is low enough such that it may be overcome by implementing Forward Error 
Correction (FEC) protocols. Including FEC means that our 6-bit photon must be partitioned to contain error correction, key, and message bits. FEC requires redundancy to detect and 
correct errors. Unfortunately, this redundancy can aid Eve in her attempt to unlock the message. As shown in \cite{lupo2015quantumArXiv}, Alice and Bob can implement error correction while maintaining security, at the cost of a higher key consumption rate, where the extra key is used to cover up the redundancy in the error correction code.  

\begin{figure}[h]
\centering
\includegraphics[width=.48\textwidth]{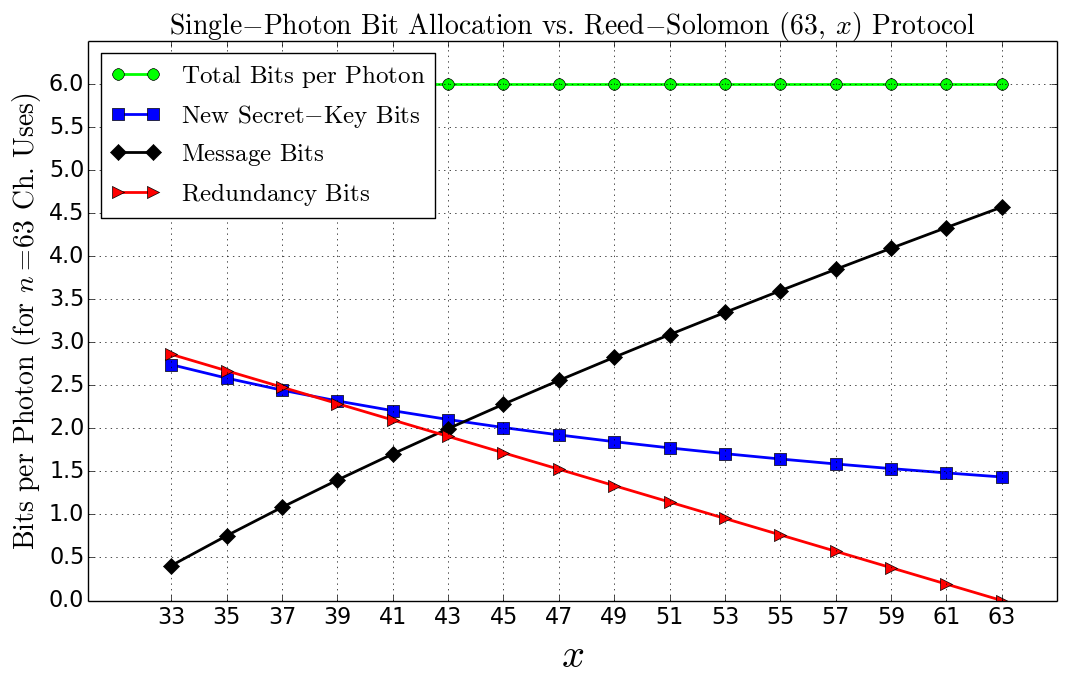}
\caption{\textbf{Single-Photon Bit Allocation (for $n = 63$ channel uses): } The plot above lists how many secret-key bits (blue squares), FEC redundancy bits (red triangles), and message bits (black diamonds) are encoded within a single 6-bit photon as a function of Reed-Solomon (63,\,$x$) codes. The new secret-key plot shows the necessary allocation of bits to replenish the consumed key. 
Any FEC redundancy requires the consumption of more secret key. Moving from $n = 63$ channel uses to $n = 126$ or higher will result in less secret-key consumption per photon and larger message-bit capacities. Messages bits may be allocated to either additional key or secret messages for direct communication.} 
\label{fig:bit} 
\end{figure}

We used Reed-Solomon error correction codes \cite{reedsolomon1960polynomial}. Reed-Solomon codes detect and correct 
errors on symbols (where each symbol is represented by several bits) rather than correcting errors on individual bits. Thus, Reed-Solomon codes treat all bit errors on a 
single symbol as a single error. This is particularly advantageous to our experiment where a 6-bit symbol is encoded by a single photon. Reed-Solomon codes also transmit in packets of symbols, or `blocks', with the largest block length being $2^s-1$ symbols for an $s$-bit alphabet. Hence, transmissions with 63 symbols per packet correspond to the most efficient use of Reed-Solomon codes for our 6-bit alphabet. We implemented Reed-Solomon (63,\,$x$) protocols that 
encode in packets of 63 symbols, or 63 photons as is our case, where $x < 63$ is the number of symbols in the original data packet containing a message and new secret key. Of those 63 symbols, $63-x$ symbols encode the redundant 
information. A Reed-Solomon (63,\,$x$) code can correct up to $(63-x)/2$ symbol errors.  
The fractional redundancy of the useful information 
is $(63-x)/x$. To cover up this redundancy, the secret key is scaled to be $1+(63-x)/x$ times larger. This leaves less information that we can allocate 
towards an encrypted message.  
Because the key rate is 
larger, the code book length will be exponentially larger. The number of bits allocated to message, redundancy, and new secret key operate as a function of the Reed-Solomon (63,\,$x$) code used. Because of this, Fig. \ref{fig:bit} presents the bit-allocation per photon as a function of the Reed-Solomon (63,\,$x$) code implemented for $n=63$ uses of the channel. Note that the allocated new secret-key bits within Fig. \ref{fig:bit} are equal to the number of consumed-key bits used to secure the transmission of the 6-bit photon. 

Ideally, a single photon source will transmit a single photon on demand. Such sources are still in development, forcing us to rely upon heralding with SPDC. SPDC dictates 
that numerous down-converted events will take place randomly per SLM setting. While alternate versions of a quantum enigma machine may allow the transmission of more than one photon per channel use (as in \cite{PhysRevA.92.062312}), this version of a quantum enigma machine should have an accurate measure of the losses and only allow one photon transmission per channel use. These two points are, perhaps, the greatest limitations of a quantum enigma machine. However, the strength of a quantum enigma machine is not based on novel hardware, nor a particularly clever protocol. The strength of a quantum enigma machine resides in the use of a weaker security definition in comparison to QKD. Therefore, many of the techniques used in QKD to cope with photon splitting an other attacks can be equally applied to a quantum enigma machine.

To demonstrate the capability of this system if we were using an ideal single-photon source, we recorded only the first heralded event per SLM setting and neglected the rest. Figure \ref{fig:FEC} presents the success rates for different Reed-Solomon error correction protocols as a function of the number of key and message symbols $x$. When only considering our first-count heralded events, we transmitted 420 packets of 63 symbols with varying error correction capabilities. Reducing the available capacity for message symbols allowed for more redundancy symbols with higher transmission success rates. In our analysis, a single incorrect symbol in the corrected packet corresponded to a complete failure of the entire packet. Figure \ref{fig:FEC} also plots the available message capacity per photon as a function of $x$ and the number of channel uses $n$. Because the Reed-Solomon codes require us to transmit in packets of 63 symbols, we chose to make the number of channel uses $n$ a multiple of 63. When limiting the message to 1.02 bits per photon ($x = 35$, $n = 126$) after having already allocated space to FEC and a new secret key, we achieved a 99.5\% packet success rate. While our bit-error ratio is significantly larger than telecommunications standards (typically less than $10^{-6}$ bit errors per bits transmitted \cite{BER1984}), the experiment is merely meant as a proof-of-principle demonstration of our ability to lock message and new secret keys with success rates approaching $100\%$. Clearly, errors in the transmission will destroy Bob's 
ability to decode future messages. Additional redundancy bits or different forms of error reduction and error correction will be required to make a quantum enigma machine a practical form of secure communication.

\begin{figure}[h] 
\centering
\includegraphics[width=.48\textwidth]{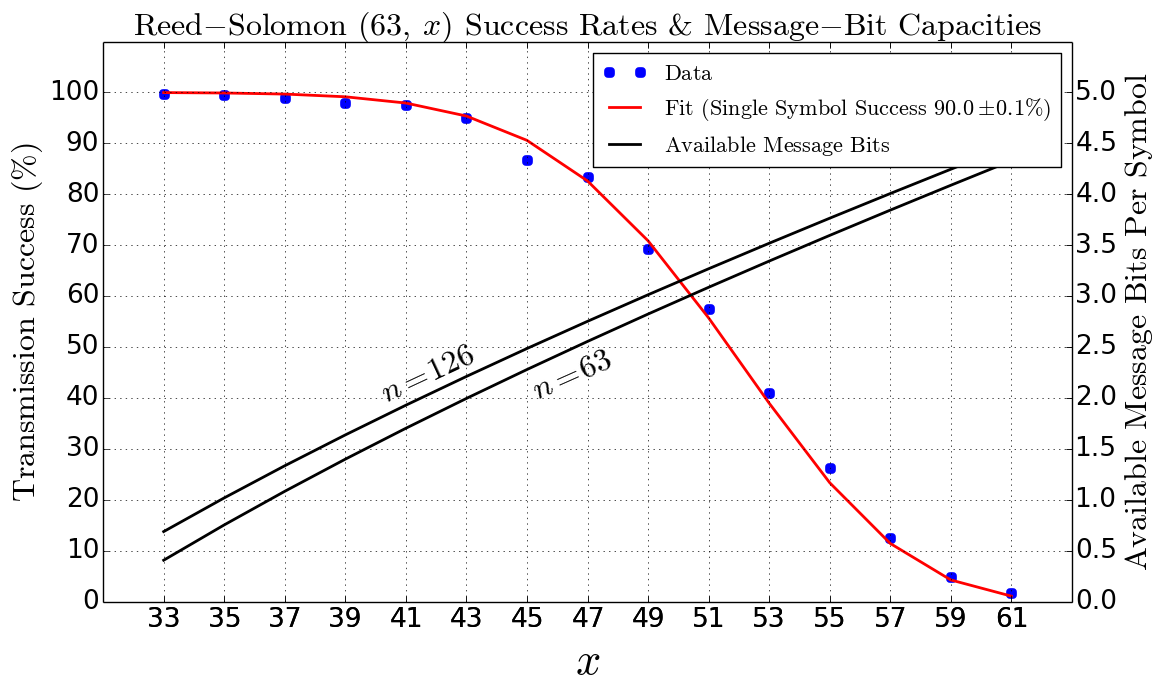}
\caption{\textbf{Reed-Solomon Error Correction Success Rates: } The success rates for a Reed-Solomon (63, $x$) code are presented, where $x$ is the number of 6-bit symbols containing the new key and message and 63 is the total number of packet symbols after including redundancy. This data was obtained after transmitting packets of 63 photons 420 times. The plot also depicts the available capacity for message bits per photon as a function of $x$ and the number of channel uses $n$.} 
\label{fig:FEC} 
\end{figure}

\section{Conclusion}

In conclusion, we demonstrated the phenomenon of quantum data locking with a proof-of-principle experiment where we applied an information-theoretic secure encryption scheme to
lock 6 bits per photon while using less than 6 bits per photon of secret key. To demonstrate the feasibility of locking both messages and new secret keys, we securely applied a Reed-Solomon (63,35) protocol and transmitted 420 packets of 63 photons (where each photon was a 6 bit symbol containing 1 bit of message, 2.3 bits of new secret key, and 2.7 bits of redundancy) with a success rate of $99.5\%$.

Although QDL has been known for about ten years, only a handful of theoretical
papers were devoted to its application to cryptography and our present
contribution presents the first experimental implementation. However, since 
the biggest difference between QDL and standard QKD is not in the hardware
but in the security definition, we believe that most of the knowledge 
accumulated in standard QKD (e.g., analysis of finite-size protocols, 
use of decoy states, measurement device independence) can be straightforwardly
transferred to QDL. 

While our implementation relies on free-space propagation, long-distance direct communication requires that QDL be tailored for fiber-optic transmission using, for example, phase modulation \cite{lupo2014robust} or the continuous-variable modulation of coherent states with homodyne detection \cite{PhysRevA.92.062312,GuhaPhysRevX,grosshans2003quantum}. To elaborate, a phase-modulation scheme could be realized by encoding information in time and scrambling that information by random phase delays. Alice would apply a time delay to a signal photon to encode information and then pass the signal photon through a positive group-velocity dispersive (GVD) medium \cite{boyd2003nonlinear} to coherently spread the photon over $d$ time bins. Alice would then apply random phases, specified by a private key and a public code book, to each time bin to scramble the time signature. To recover the timing information, Bob would apply the corresponding phase delays, specified by the public code book, to each time bin and then pass the resulting photon through an appropriate negative GVD medium.

Our free-space experiment 
presents a fundamentally simple implementation that is a stepping stone to efficient information-theoretic secure direct communication. This quantum enigma machine marks one of the first experiments to implement QDL, circumventing Shannon's standard, in an attempt to remain information-theoretically secure under the assumption that both legitimate receivers and adversaries have quantum memories with limited storage times.

\section*{Funding Information}

This work was sponsored by DARPA Grant No. W31P4Q-12-1-0015 and AFOSR Grant No. FA9550-13-1-0019.

\section*{Acknowledgments}
$\star$ Contribution of NIST, an agency of the U.S. government, not subject to copyright \\

The authors thank James Schneeloch and Samuel Knarr for editing and useful input.

\newpage

\appendix*

\section{Secret-key length}

Quantum Data Locking provides a secure method of information transfer such that two legitimate parties, Alice and Bob, can effectively communicate encrypted information via quantum states over a $d$ dimension quantum channel while limiting an eavesdropper, Eve, to an accessible information of
\begin{equation}
I_{\text{acc}} \leq \mathcal{O}\left(\epsilon \log_2 \left(d^n\right )\right)
\label{eq:Iacc}
\end{equation}
bits when using a quantum channel $n$ times where $\epsilon$ grows exponentially small with channel use, i.e. $\epsilon = 2^{-n^c}$ for $c < 1$. The foundation for our theory is elegantly outlined in \cite{lupo2015quantumArXiv}. Our experiment encodes a 6 bit symbol $x$ onto a single photon state $\ket{x_c}$ for $c = 1,2,...,\,64$. The transmission of a single photon counts as a single use of the quantum channel. Hence, for Alice to encode one of $M = 64^n$ messages within $n$ photons, she must prepare the quantum state 
\begin{equation}
\ket{\psi_c} = \bigotimes\limits_{j=1}^n\ket{x_{j,c}}
\end{equation}
and encodes each photon with a unitary transformation $U^{(s)}$ for $s = 1,2,...,K_n$. The resulting unitary operating on the $n$ photon state is
\begin{equation}
U^{(s)} = \bigotimes\limits_{j=1}^n U_j^{(s)}.
\end{equation}
The $n$ photon encrypted quantum state is then
\begin{equation}
\ket{\psi_{c}^{(s)}} = U^{(s)}\ket{\psi_c} = \bigotimes\limits_{j=1}^n U_j^{(s)}\ket{x_{j,c}}.
\end{equation} 

According to \cite{lupo2015quantumArXiv}, Eqn. \ref{eq:Iacc} is satisfied provided the number of scrambling unitaries satisfies 
\begin{equation}
K_n \geqslant \max
\begin{cases} 2 \left(\frac{2d}{d+1}\right)^n\left(\frac{1}{\epsilon^2} \ln M +\frac{2}{\epsilon^3}\ln\frac{5}{\epsilon}\right) \\[1em]
\frac{d^n}{M}\frac{4 \ln 2 \, \ln d^n}{\epsilon^2}\end{cases}
\label{eq:KeySize}
\end{equation}
where the maximum is taken over the two equations, $M = 64^n$, and $\epsilon = 2^{-n^{c}}$ for $c<1$.

The only unknown variable within Eqns. \ref{eq:Iacc} and \ref{eq:KeySize} is the dimension of the communication channel $d$. Because our quantum enigma machine relies on free-space propagation that employs a lenslet array immediately before the detector, the channel dimension can be estimated in two ways. 

\subsection{Practical estimate of $d$}

The first estimate of $d$ follows a practical line of reasoning that considers only the spatial modes that can be measured after propagating through the lenslet array. While there are numerous spatial modes that may exist within the plane of Bob and Eve's detector, only those modes that overlap the $8\!\times\!8$ lenslet array will be mapped to the $8\!\times\!8$ nanowire array. This means there are only 64 possible \emph{distinguishable} modes. All other spatial modes have $100\%$ loss and cannot be used to transmit information. If we limit Eve to the same constraints as Bob, then Eve can only measure the modes that overlap with her $8\!\times\!8$ lenslet array. Since each lens within the lenslet maps all light impinging on them to one unique nanowire, all detectable transmissions must exist within a space of $d = 64$ dimensions.

\begin{figure}
\centering
\includegraphics[width=.35\textwidth]{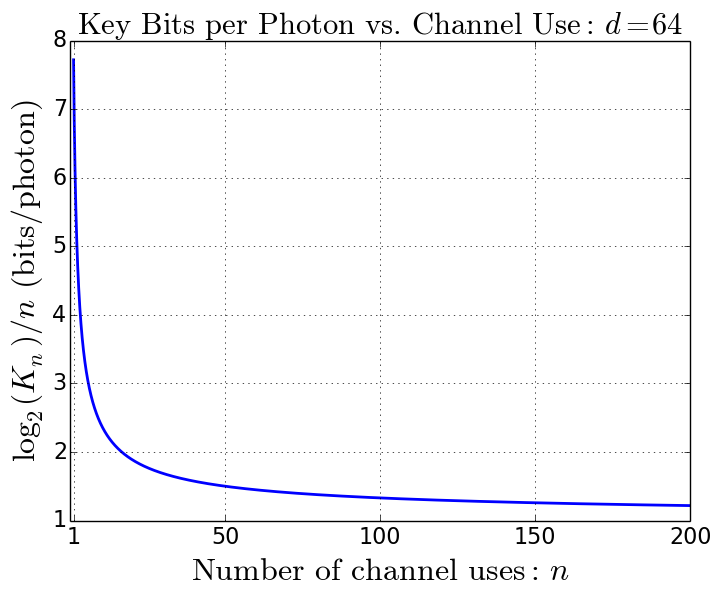}
\caption{\textbf{Secret-key consumption rate: }Letting $d = 64$, Eqn. \ref{eq:KeySize} is used to calculate the amount of key bits per photon (assuming one photon per channel use). QDL is only possible if $(\log_2K_n)/n < 6$; we see that QDL is possible for most values of $n$.}
\label{fig:KeyRate64}
\end{figure}

Letting $d = 64$, $M = 64^n$, and $\epsilon = 2^{-\sqrt{n}}$, Eqn. \ref{eq:KeySize} is used to derive the necessary amount of key per photon and is  plotted within Fig. \ref{fig:KeyRate64}.

In addition to the standard QDL protocol, we also apply Reed-Solomon forward error correction (FEC) codes using symbols of 6 bits per photon. The optimal block length for a Reed-Solomon code with a 6-bit alphabet is 63 symbols within a block. Thus, it is reasonable to set the number of channel uses to be a multiple of 63. In our analysis, we consider two cases: $n = 63$ and $n = 126$.

From Fig. \ref{fig:KeyRate64}, we show that $(\log_2 K_{63})/63 \geqslant 1.434$ bits per photon and $(\log_2 K_{126})/126 \geqslant 1.287$ bits per photon is enough to limit $I_{\text{acc}}$ to only a few bits compared to having transmitted 378 bits and 756 bits respectively. The resulting message bit capacities and Reed Solomon success rates are shown within the main article.

\subsection{Conservative estimate of $d$}

A more conservative estimate of the channel dimension $d$ is obtained by considering all of the possible distinguishable spatial modes that can exist after passing through an infinite dimensional lenslet array, as opposed to only those that can be measured. The lenslet array effectively maps a high-dimensional spatial pattern to a lower dimensional one. 
$d$ is obtained by dividing the average distribution in a plane of Eve's detector by the area of a circular lens within an infinite dimensional lenslet array. This effectively overestimates the dimension because a circular lens was used within the experiment and areas between the circular lenses correspond to `dead-space' where light cannot be mapped to a detector. The choice of a circular lens area, instead of a square area, results in a larger $d$ and an over-secure key rate.  

Eve's average distribution was obtained by modeling the propagation of light through the system. Our experimental setup used a $512\!\times\!512$ pixel spatial light modulator (SLM) to both scramble and encode linear phases onto a single photon projected from a single-mode optical fiber. The transverse profile of our single-photon state takes the form
\begin{equation}
\ket{x_c} = 
\frac{1}{\mathcal{N}}\iint\limits_{-\infty}^{\infty}\,
\text{e}^{\text{i}\left(\phi_{x,c}x+\phi_{y,c}y\right)}
\text{e}^{\left\{-\frac{x^2+y^2}{4\sigma^2}\right\}} \,dx\,dy\,\ket{x,y}
\end{equation}
where $\mathcal{N}$ is a normalization constant and $\phi_{x,c}$ and $\phi_{y,c}$ are the necessary linear phases to encode a message with specifier $c = 1,2,...,64$ into the transverse profile of a single photon state whose probability distribution follows a Gaussian with a standard deviation of $\sigma$. Using the SLM, Alice simultaneously applies the linear phase encoding the message and encrypts each photon wavefront with a scrambling unitary operation $U_j^{(s)}$. The unitary operations are composed of random binary phase patterns containing the values 0 and $\pi$. While each scrambling unitary consists of $512\!\times\!512$ SLM pixels, superpixels composed of $4\!\times\!4$ SLM pixels were used to generate $128\!\times\!128$ superpixel resolution random binary phase patterns. The lower resolution binary patterns aided in the alignment of Alice's SLM to the image plane on Bob's SLM. Each unitary has the form
\begin{equation}
U_j^{(s)} = \sum\limits_x\sum\limits_y\text{e}^{\text{i}\theta_k(x,y)}\ket{x,y}\bra{x,y}
\end{equation}
such that $\theta_k(x,y)$ is a random binary variable (0 or $\pi$) chosen from a uniform distribution.
This probability amplitude is propagated to the focal point of a lens using a Fourier transform defined by
\begin{align}
\begin{split}
G(f_x,f_y) & = \iint\limits_{-\infty}^{\infty}\,g(x,y)\text{e}^{-2\pi\text{i}\left(f_x x+ f_y y\right)} dx \, dy \\
& = \mathcal{F}\left[g(x,y)\right].
\end{split}
\end{align}
To obtain the probability distribution in either Bob's or Eve's focal plane ($P_{\text{Bob}}$ or $P_{\text{Eve}}$), we need only consider the following: 
\begin{align}
 P_{\text{Bob}} &= \frac{1}{\mathcal{N}} \left|\mathcal{F}\left[
\text{e}^{\left\{-\frac{x^2+y^2}{4\sigma^2}\right\}}\text{e}^{\text{i}\left(\phi_{x,c}x+\phi_{y,c}y\right)}\right]\right|^2
\label{eq:PBob}
 \\
 P_{\text{Eve}} &=\frac{1}{\mathcal{N}} \left|\mathcal{F}\left[
\sum\limits_{x,y}\text{e}^{\left\{-\frac{x^2+y^2}{4\sigma^2}\right\}}\text{e}^{\text{i}\left(\phi_{x,c}x+\phi_{y,c}y+\theta_k(x,y)\right)}\right]\right|^2
\label{eq:PEve}
\end{align}
where $\mathcal{N}$ is the necessary normalization constant to form a probability distribution.

While these probability distributions are infinite dimensional, the actual detector system within the experiment contained a lenslet array placed one focal length in front of the nanowire array. The lenslet array mapped the area of a single lens onto a single nanowire. If we extend the lenslet array to operate on the entire distribution, Eve's distribution is then effectively discretized by the area of a lens. The resulting dimension $d$ is then the ratio of Eve's average distribution area to the area of a single lens within the lenslet array.

The actual Fourier transform probability distributions, residing within the detector's cryostat, were not measured. However, we can infer both the area of Eve's distribution and the diameter of lenses within the lenslet array by knowing the initial beam profile on Alice's detector and by using the 64 nanowire phase settings within the experiment. Magnification effects, used to focus the light down onto a single nanowire, are not required within the calculation of $d$. The magnification is not required because it will inevitably cancel when calculating the ratio of Eve's distribution area to the lens area. We chose to perform this calculation numerically.

\begin{figure}
\centering
\includegraphics[width=.35\textwidth]{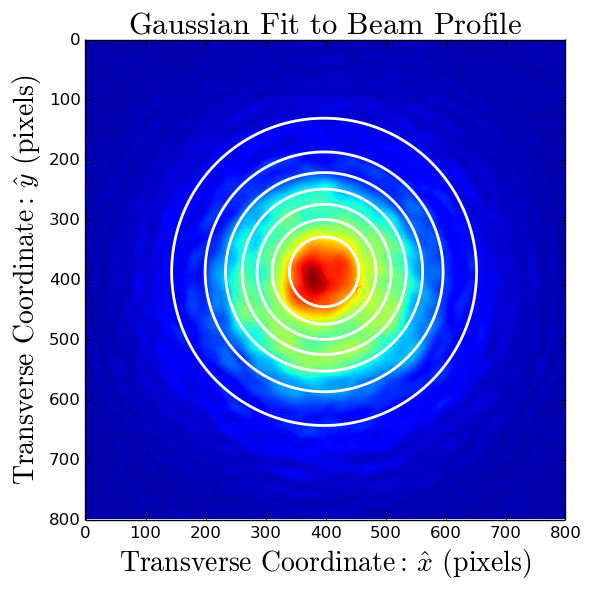}
\caption{\textbf{Gaussian fit to beam profile: } A Gaussian function was fit to the original beam profile seen within the experiment. This fit was used to find the original beam profile on Alice's SLM for a numerical simulation.}
\label{fig:GaussFit}
\end{figure}

The lens area was calculated using experimental settings. Linear phase settings were optimized within the experiment to have focal points centered on each nanowire. Hence, the linear phases should be an accurate measure of the lens spacing within the lenslet array. Using these linear phase settings, we Fourier transformed them to find the lens centers. The lens separations were found in units of pixels (relative to the resolution of the numerical Fourier transform). The Fourier transform was zero padded to resolve fine details -- one pixel on Alice's SLM was mapped into 9 Fourier transform pixels.

To obtain the average area of Eve's distribution, the relative area of the beam profile on a $512\!\times\!512$ pixel SLM needed to be calculated to perform a numerical simulation. The relative area of the initial beam on the SLM (in units of SLM pixels) was obtained by first measuring the collimated beam profile found in the actual experiment. Using the asphercal lens from the experiment to collimate light emitted from a single-mode optical fiber, the beam profile was imaged with a camera and a Gaussian function was then fit to this profile, as shown in Fig. \ref{fig:GaussFit}. 
Knowing the pixel sizes for the camera and SLM, we calculated the necessary standard deviation of a Gaussian beam such that its profile would fill the same relative area on our SLM for the numerical simulation.




\begin{figure}
\centering
\includegraphics[width=.35\textwidth]{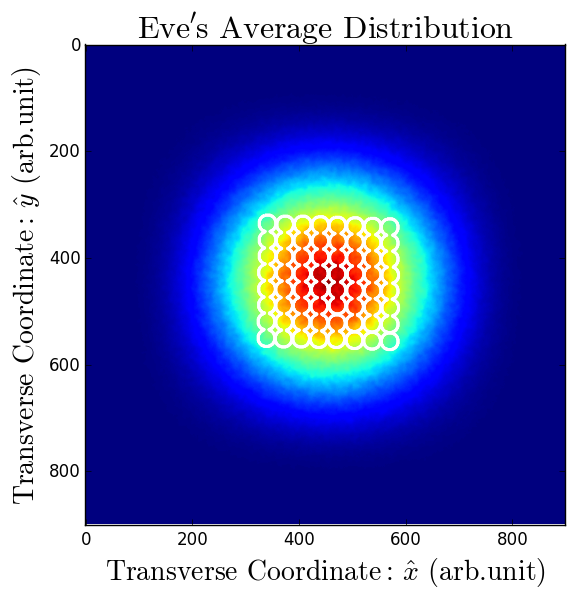}
\caption{\textbf{Eve's average distribution: } This distribution was obtained by averaging over each of the 64 scrambled messages 300 times whereby a different random phase pattern was used for each scrambling operation. The lenslet array, calculated from experimental parameters, is outline in white.}
\label{fig:EveDistrib}
\end{figure}

Eve's probability distribution was calculated by averaging over each of the 64 possible messages. Each message was scrambled 300 times by different random $128\!\times\!128$ binary phase patterns composed of 0 and $\pi$ phase shifts. After averaging over each of the 64 messages, the distribution was thresholded, keeping only those values within a $4\sigma$ radius of the resulting zero-order distribution. The average distribution can be seen in Fig. \ref{fig:EveDistrib}. Notice the size of this distribution in comparison to the lenslet size. To approximate the dimension of the communication channel, the number of non-zero pixels within a $4\sigma$ radius of Eve's average distribution was divided by the lenslet area. This simulation was done 10 times, while averaging over 300 scrambling operations for each of the 64 messages, to arrive at a dimension $d = 644.8 \pm 0.8$.

Within the actual experiment, the lenslet array was placed approximately 1 focal length away from the nanowire array. Hence, Bob and Eve would not have placed their lenslet in the focus of their Fourier transform lens. The degree of misfocus can be accounted for by introducing a misfocus operator into the propagation equations taking the form of a quadratic phase factor associated with a spherical lens. We describe the misfocus by an operator of the form
\begin{equation}
F_{\text{error}} = \sum\limits_x\sum\limits_y\text{e}^{i\alpha\left(x^2+y^2\right)}\ket{x,y}\bra{x,y}
\end{equation}
where $\alpha \ll 1$. Note, this quadratic phase must be applied before the Fourier transform. 

\begin{figure}[h]
\centering
\includegraphics[width=.4\textwidth]{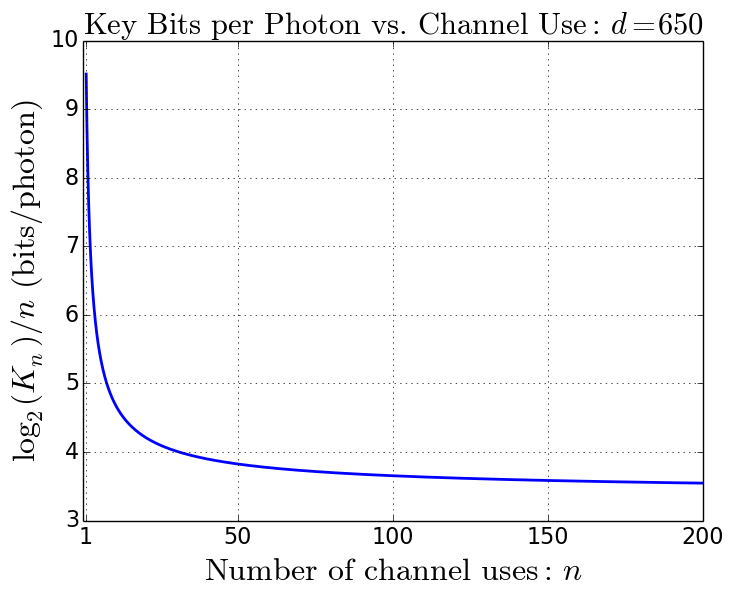}
\caption{\textbf{Secret-key consumption rate: }Letting $d = 650$, Eqn. \ref{eq:KeySize} is used to calculate the amount of key bits per photon (assuming one photon per channel use). QDL is only possible if $(\log_2K_n)/n < 6$; we see that QDL is possible for most values of $n$. For $n \gtrsim 50$, a key size just under 4 bits per photon is sufficient for QDL and has the potential to enable the use of error correcting codes.}
\label{fig:KeyRate650}
\end{figure}

Repeating the previous simulation with any reasonable degree of misfocus (up to a maximum crosstalk error of $50\%$ -- significantly larger than in the actual experiment), we always obtain a smaller channel dimension $d$ compared to that of the previous simulation with no misfocus. Hence, the largest possible dimension is rounded up slightly to $d = 650$ to err on the side of security. Overestimating slightly requires a longer key, meaning we can transmit fewer message bits over $n$ uses of the channel. However, sacrificing communication bits within reason is a safer alternative to not limiting $I_\text{acc}$ sufficiently.

\begin{figure}[h]
\centering
\includegraphics[width=.48\textwidth]{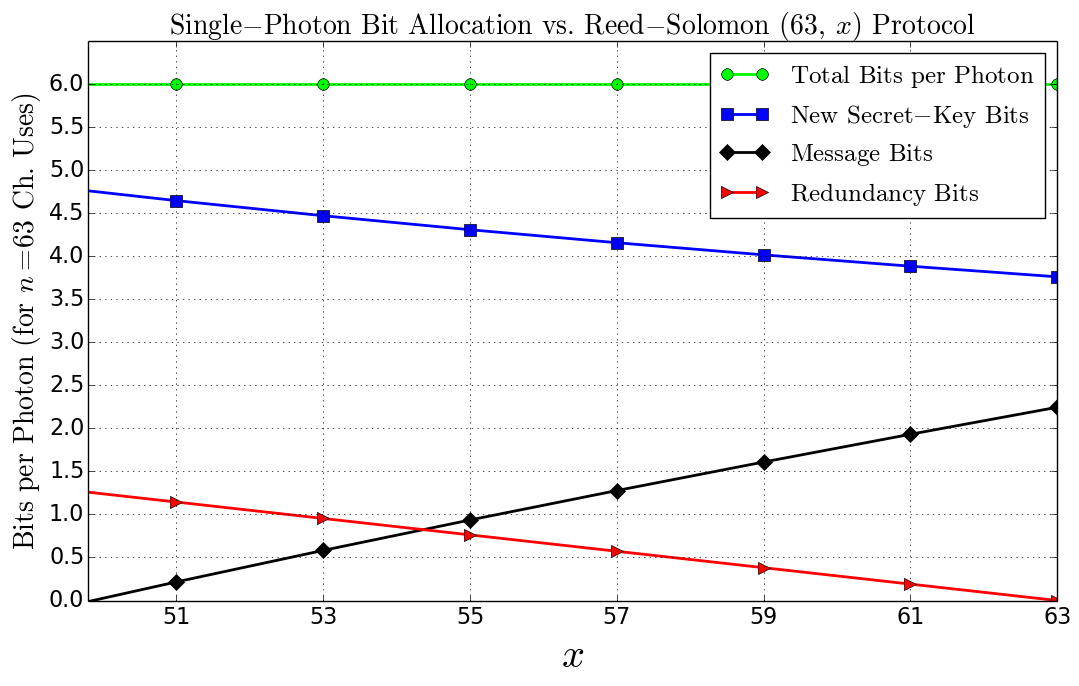}
\caption{\textbf{Single-photon bit allocation (for $n = 63$ channel uses): } The plot above lists how many secret-key bits (blue squares), FEC redundancy bits (red triangles), and message bits (black diamonds) are encoded within a single 6-bit photon as a function of Reed-Solomon (63,\,$x$) codes for a channel dimension $d = 650$. The new secret-key plot shows the necessary allocation of bits to replenish the consumed key. 
Any FEC redundancy requires the consumption of more secret key. Moving from $n = 63$  to higher values will result in less secret-key consumption per photon and larger message-bit capacities as the key asymptotically approaches a constant. The Reed-Solomon (63,\,51) code holds the largest Reed-Solomon error correcting capability for $d =650$ and $n = 63$.} 
\label{fig:bitcap} 
\end{figure}

Calculating the necessary key rates for $n = 63$ and $n = 126$ while letting $d = 650$, $M = 64^n$, and $\epsilon = 2^{-\sqrt{n}}$, Fig. \ref{fig:KeyRate650} plots $(\log_2K_n)/n$ to arrive at the necessary amount of key bits per photon to secure the transmission. As long as $(\log_2 K_n)/n < 6$, QDL is possible for given $n$. From the plot, we show that $(\log_2 K_{63})/63 \geqslant 3.757$ bits per photon and $(\log_2 K_{126})/126 \geqslant 3.611$ bits per photon is enough to limit $I_{\text{acc}}$ to only a few bits compared to having transmitted 378 bits and 756 bits respectively. While the two different values of $(\log_2K_n)/n$ for different $n$ calculated here vary only slightly, the option to move from $n = 63$ to $n = 126$ or higher may have a profound impact on the available message bits once the redundancy of the Reed-Solomon codes has been accounted for, i.e. the key bits are scaled linearly with the redundancy corresponding to an exponential change in the code book size. Because the amount of allocated message, redundancy, and key bits depend on the Reed-Solomon (63,\,$x$) code used, Fig. \ref{fig:bitcap} plots bit allocation within a 6-bit photon for $n = 63$ uses of the channel as a function of $x$. Again, note that the new secret-key bit plot is equal to the consumed key bits necessary to secure the transmission of the 6-bit photon for $n = 63$ uses of the channel. Also notice that QDL with FEC becomes impossible once the message-bit capacity becomes $0$. Allocating a larger number of channel uses $n > 63$ will be required to reduce the key length and allow for more message and redundancy bits.

\begin{figure}
\centering
\includegraphics[width=.48\textwidth]{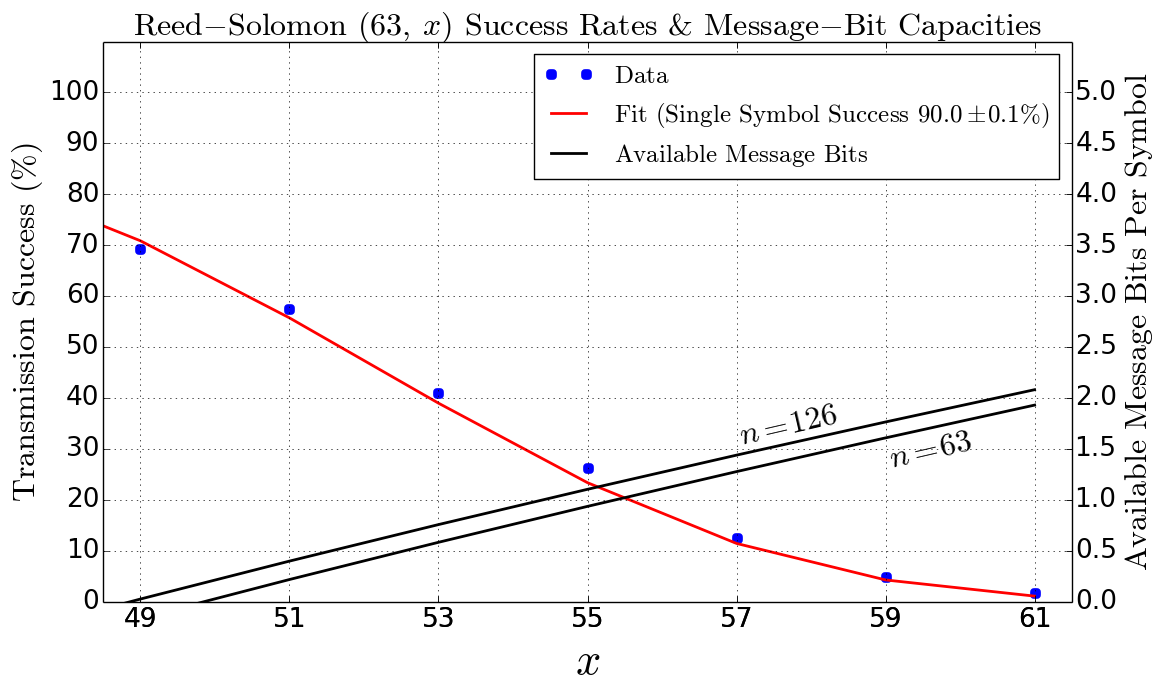}
\caption{\textbf{Reed-Solomon error correction success rates: } The success rates for a Reed-Solomon (63,$x$) code are presented, where $x$ is the number of 6-bit symbols containing the new key and message and 63 is the total number of packet symbols after including redundancy. This data was obtained after transmitting packets of 63 photons 420 times. The plot also depicts the available capacity for message bits per photon as a function of both $x$ and the number of channel uses $n$ after allocating the necessary bits to refresh a secret encryption key obtained from Eqn. \ref{eq:KeySize} with $d = 650$. From the plot, we see that less key is required for $n = 126$ resulting in a higher message capacity.} 
\label{fig:FEC} 
\end{figure}

The corresponding impact on the available message bits per photon are shown in Fig. \ref{fig:FEC}. QDL is still possible, but the Reed-Solomon code is not sufficient to correct the majority of errors while providing message transmission capabilities on top of new secret-key distribution.

Moving forward, in order to utilize QDL systems operating over high-dimensional quantum channels, either detector systems must improve to make the available message space larger (i.e., increasing the efficiency of key use), or more efficient FEC protocols must be implemented to increase the error correcting capability with fewer redundancy bits.

\section{Additional experimental details}

Single photon pairs were generated through type-I degenerate Spontaneous Parametric Down Conversion (SPDC) using a nonlinear bismuth triborate (BiBO) crystal. Ultraviolet (UV) laser light at 325 nm first passed through a notch filter to clean the pump beam. The UV light was later filtered out with a pump filter while passing the 650 nm down-converted signal and herald photons. The signal and herald photons were then separated into different paths by a 50/50 beamsplitter and coupled into single-mode fibers. The signal photon was launched, collimated, and then prepared in a polarization state to be phase modulated by a polarization sensitive Meadowlark $512\!\times \!512$ pixel liquid crystal reflective type SLM. Alice modulated her state with both a linear phase and a scrambling phase mask. The linear phase was chosen from a set of 64 linear phases that directed photons to a specific nanowire within the nanowire array. Hence, linear phases encoded 6 bits of information. Imaging optics within Eve's domain mapped the scrambled state from Alice's SLM to Bob's SLM. Bob phase modulated his state with the inverse scrambling unitary operation before focusing the light onto his $8\!\times\! 8$ single-photon-detecting nanowire array \cite{allman2015near}. The nanowires were cryogenically cooled to 0.8 K and were covered with an $8\!\times\! 8$ lenslet array. Each lenslet was 150 $\mu$m in diameter and focused light onto each nanowire. If Bob applied the correct inverse scrambling unitary operation, only a single nanowire should receive photons. Alice's PerkinElmer Avalanche Photodiode (APD) was used to herald the presence of a signal photon on Bob's detector using high speed correlating electronics. We were unable to guarantee that a single photon was transmitted through the channel for each SLM setting (corresponding to one use of the channel). To demonstrate the potential of this implementation once single-photon sources become commercially available, we only recorded the first heralded event per SLM setting in our message-passing data analysis with FEC.

Because of the losses within the system (including detectors, SLM's, optics, and polarizers outside of Eve's domain), we assumed a weak-locking capacity \cite{GuhaPhysRevX} where Eve does not have access to Alice's SLM. While the 4F imaging system within Eve's domain contained anti-reflection coatings, we neglected the losses that could be due to Eve. However, the experiment is meant as a proof-of-principle example because we could not alleviate the losses due to all other optical components and equipment. If these losses can be accurately accounted for, it has been shown in \cite{lloyd2013quantum} that security according to a strong locking capacity, where Eve has direct access to a noiseless state from Alice's SLM \cite{GuhaPhysRevX}, can still be obtained when dealing with an arbitrary amount of loss by making both the code book and the dimension of Alice's system larger. 

\bibliography{enigma}

\end{document}